\documentclass[prb,showpacs,twocolumn,amssymb,amsmath,nobibnotes]{revtex4}

\usepackage{graphicx,color}% Include figure files
\usepackage{dcolumn}% Align table columns on decimal point
\usepackage{bm}% bold math

\begin{document}

\title{Graphene on metal surface: gap opening and $n$-doping}% Force line breaks with \\

\author{Y. H. Lu}
\affiliation{Department of Physics, Zhejiang University, Hangzhou, China}
\author{P. M. He}
\email{phypmhe@dial.zju.edu.cn}
\affiliation{Department of Physics, Zhejiang University, Hangzhou, China}
\author{Y. P. Feng}
\email{phyfyp@nus.edu.sg}
\affiliation{Department of Physics, National University of Singapore, 2 Science Drive 3, Singapore, 117542}

\begin{abstract}
Graphene grown on metal surface, Cu(111), with a boron nitride(BN) buffer layer
is studied for the first time. Our first-principles calculations reveal that
charge is transferred from the copper substrate to graphene through the BN
buffer layer which results in a $n$-doped graphene in the absence of a gate
voltage. More importantly, a gap of 0.2 eV which is comparable to that of a
typical narrow gap semicondutor opens just 0.5 eV below the Fermi-level at the
Dirac point. The Fermi-level can be easily shifted inside this gap to make
graphene a semiconductor which is crucial for graphene-based electronic
devices. A graphene based $p$-$n$ junction can be realized with graphene
eptaxially grown on metal surface.

\end{abstract}

\maketitle

Graphene, first isolated by Novoselov\cite{Novo}, have attracted a lot of
research interest recently because of its intriguing physics for fundamental
studies and its potential applications for the next generation electronic
devices, such as novel sensors\cite{sensor} and post-silicon
electronics.\cite{e-divice1,e-divice2,e-divice3,e-divice4} As a 2D crystal,
graphene was found not only to be continuous but to exhibit high crystal
quality,\cite{Novo,crystal1,crystal2} in which charge carriers can travel
thousands of interatomic distances without being scattered.\cite{charge1,
charge2, charge3} In addition, due to its special honeycomb structure, band
structure of graphene exhibits two  intersecting bands at two inequivalent K
points in the reciprocal space and its low energy excitations are mass-less
Dirac fermions near these K points because of its linear (photon-like)
energy-momentum dispersion relationship. This results in very high electron
mobility in graphene which can be further improved significantly, even up to
$\approx 10^5$ cm$^2$/V$\cdot$s. It also allows controls of carrier type
(electron-like or hole-like) and density by electric-field\cite{Novo}. These
are different from conventional doping of semiconductors, for example via ion
implantation. With such features, graphene is promising in many applications,
such as bipolar devices which comprises junctions between hole-like and
electron-like regions, and \textit{p-n} junctions which can be configured by
gate-voltage within a single atomic layer.\cite{gate1,gate2,gate3} Graphene
based devices  can be expected to have much more advantages than silicon-based
devices.

At present, there are mainly two methods of producing graphene samples. In the
first, an almost freestanding graphene is produced by mechanically splitting
off bulk graphite crystals and depositing it onto a SiO$_2$/Si
substrate.\cite{SiO2} Although this method is very convenient, it is difficult
to produce high quality graphene structures. In the second method, an ultrathin
graphene layer is formed by vacuum graphitization through silicon depletion
from a SiC surface.\cite{SiC} The large lattice mismatch between SiC substrate
and graphene results in graphene produced using this method being fragile and
containing many defects.\cite{SiC-defect} It is thus desirable to find a way of
fabricating graphene which can overcome the above disadvantages. Pristine
graphene is a zero-gap semiconductor because its Fermi level exactly crosses
the Dirac point. An energy gap is required for many practical applications.

A gap can be induced in graphene by a weak interaction with the substrate when it
is grown on a suitable substrate.\cite{gap} However, choosing the right
substrate is not trivial. If the interaction is strong, the
advantages of graphene will be lost which is the case when a single layer of graphene grown on
SiC.\cite{cal-GronSiC} On the other hand, a too weak interaction between graphene and the substrate is not able to change the property of graphene and to open a
gap.\cite{GrhBN}
Attempts were made to grow graphene directly on metal substrate,\cite{GronNi,moire}which provides an interaction stronger than van der Waals interaction (such
as graphene grown on h-BN substrate\cite{GrhBN}) but weaker than covalent interaction (such as that between graphene and a SiC
substrate\cite{cal-GronSiC}), which indeed led to opening of band gap.
In addition, large area of flat single-crystal metal surface is
easy to obtain experimentally. Therefore, metal substrate can be useful in
graphene-based electronic-devices. The only drawback is that the energy bands
and the Dirac cone of graphene are submerged in the electronic sea of metal and the unique property of graphene cannot be utilized when
it is grown directly on metal substrate. It is thus important to find proper
substrate to overcome this disadvantage. In this work, we investigate
graphene grown on Cu(111) substrate with a boron nitride (BN) buffer layer
between them. The experimental lattice constant of Cu(111) surface is 2.55 \AA\
which is very close to those of the BN sheet (2.50 \AA) and graphene (2.46\AA).
The slight mismatch is acceptable since the C-C bond and B-N bond are very
strong. It is noted that in the growth of graphene directly on SiC, the excess
carbon atoms self-organize to form the honeycomb structure which overwhelms the
covalent bonding with substrate, even through the lattice mismatch between the
graphene layer and the SiC substrate is as high as 8\%.\cite{SiC-defect} Thin
carbon layer consisting of sp$^2$-hybridized carbon modification in the form of
graphene stacks over copper nanoparticles have been produced.\cite{Cuparticle}
Heteroepitaxial graphene/h-BN double layer has also been realized on some metal
substrate by chemical vapor deposition(CVD).\cite{GrBNonNi} It can be expected
that including a BN buffer layer will improve the stability of the
heteroepitaxial graphene/h-BN structure on Cu(111) surface. Furthermore, the
special electronic structure of this system ensures that it will be more
suitable for applications in electronic devices, which will be further
discussed below.

First-principles calculations based on the density functional theory were
performed using the  VASP code\cite{vasp1,vasp2,vasp3,vasp4} The projector
augmented wave pseudopotentials\cite{paw}were used for electron-ion
interactions while the local density approximation (LDA) was used for
exchange-correlation function. A special 36$\times$36$\times$1 $k$-point
sampling was used for the surface Brillouin-zone integration. The plane wave
basis set was restricted by a cutoff energy of 400 eV. The in-plane lattice
constant of the heteroepitaxial graphene/h-BN/Cu(111) structure was set to 2.49
\AA\ which is the lattice parameter of Cu(111) determined from our
first-principles calculation within LDA. In structural relaxation, five atomic
layers of Cu were used to model the substrate, with atoms in the bottom two
layers  fixed to their bulk positions and h-BN and graphene adsorbed to the top
surface. Electronic structure calculations were carried out using a ten layer
substrate model with two center  layers fixed. Graphene and BN layer are
adsorbed on both sides. A vacuum region of at least 18 \AA\ was used to avoid
interaction between the top and bottom surfaces. All structures were fully
relaxed, and the forces acting on an atom is $\le 0.03$ eV/\AA\ in the relaxed
structures.

The graphene and BN layers were placed on the $1\times 1$ unit cell of Cu(111).
There can be  different atomic arrangements for BN and graphene relative to the
substrate. First of all, the BN sheet can have six different arrangements over
Cu(111), with either B or N on the fcc or hcp site, and atop a Cu atom of the
substrate, respectively. Among these, two arrangements, with N on the atop site
while B on either the fcc or hcp site as shown in Fig.~1(a) and Fig.~1(b),
respectively (referred as structures A1 and A2 respectively in the following),
are energetically favored. They are very similar to the structure of BN sheet
on Ni(111).\cite{GrBNonNi} The energies of these two structures are almost the
same and are lower than those of other structures by more than 20 meV. The
distance between the BN sheet and the Cu(111) surface is about 2.70 \AA\ for
the A1 arrangement and 2.79 \AA\ for the A2 arrangement, which is smaller than
the distance between adjacent atomic layers in bulk h-BN (3.35 \AA) but larger
than that between a BN sheet and the Ni(111) surface.\cite{GrBNonNi} Therefore,
the strength of interaction between the BN sheet and Cu(111) can e expected to
be stronger than that between the BN layers and weaker than that with the
Ni(111) substrate.  In the relaxed structure, the B and N atom are found not in
the same plane, and the N atom which is at the atop Cu site is a slighly higher
($\sim 0.02$ \AA) than B atom which is located at the hollow site. This
rumpling is also much smaller compared to that of BN on Ni(111) substrate where
a rumpling of 0.18 \AA\ was observed experimentally. This is also an indication
of relatively weaker interaction between the BN layer and Cu(111) substrate,
compare to Ni(111).

To find the most stable structure of graphene on BN/Cu(111), self-consistent
calculations were performed for all possible geometries between the graphene
and the above two most stable BN arrangements on Cu(111). Relative to the BN
lattice, carbon atoms of the graphene can be directly over the B and N atoms,
or one carbon atom in the hollow site and the other above either a B or a N
atom. Among the six possible heteroepitaxial structures, the two atomic
arrangements A1 (Fig.1a) and A2 (Fig.1b) are energetically degenerate and are
predicted to be most stable. In both structures, one graphene sub-lattice is
over the B atom. The other sub-lattice is on the hcp hollow site of Cu(111) in
A1 while it is on the fcc hollow site of Cu(111) in A2.  The graphene floats
over the BN sheet at a height of $\sim 3.0$ \AA, slightly smaller than
inter-layer distance of graphite. This confirms that LDA, despite the lack of
long-range nonlocal correlations, produces reasonable interlayer distances in
layered van der Waals crystal, such as graphite and h-BN, for a delicate error
cancelation between exchange and correlation which underlies this apparent
performance of LDA.\cite{lda} The distance between BN sheet and Cu(111) changed
by $-0.3$ \AA\ after adsorption of graphene. However, the distance between Cu
layers remains essentially the same as that in a clean Cu(111) surface. Further
structural details are given in Table I.

Figs.~2(b) and 2(c) show the band structure of optimized atomic arrangement A2
before and after graphene adsorption, respectively. For reference, the band
structure of clean Cu(111) surface is shown in Fig.~2(a). The $d$-bands of
copper are apparent in all three band structures and are essentially unaffected
by BN and graphene due to the weak interaction with them, except at the
$\Gamma$ point near the Fermi level. After adsorption of the BN sheet, bands
characteristic of h-BN appear at about $-18$ eV in the band structure while the
occupied $p$-bands of BN appear at about 4 eV below the Fermi level at the
$\Gamma$ point. The most interesting feature is the appearance of a band in the
energy gap of Cu(111) at K about 2 eV above the Fermi level. In perfect BN
sheet, this valence band minimum is about 18.4 eV above the lowest $s$-band at
K within LDA calculation, but it is reduced to about 18.2 eV after the BN sheet
is adsorbed on Cu(111). This shift is due to surface dipole effect of Cu(111)
substrate, as indicated by the change of work function of the Cu(111) surface
from 5.25 eV to 4.83 eV after BN adsorption. Electrons redistribe in response
to the dipole effect, resulting also to the slight downward shift of some
Cu(111) bands at M and the upward shift of some bands at $\Gamma$ around the
Fermi level. Neverthless, the BN layer remain insulating since the Fermi level
does not cross any BN bands. After graphene adsorption, the Cu(111) bands at
$\Gamma$ become completely unoccupied and bands characteristic of graphene
appear in the energy range of $-18$ eV to $-4$ eV. The Dirac cone at K is
located in the energy gap of the substrate, just 0.5 eV below the Fermi level.
The linear relationship between energy and momentum is maintained. This
indicates that the graphene is $n$-doped with excess carriers induced by the
substrate. In addition, a band gap of about 0.2 eV opens at K for graphene
which is much larger than that of graphene on h-BN substrate\cite{GrhBN},
although the interfacial atomic arrangement between graphene and h-BN is the
same in both cases. As mentioned above and shown in Fig.1(b), one sublattice of
graphene is located directly above B or the hcp hollow site of Cu(111) while
the other is at the hexagonal center of h-BN honeycomb or the fcc hollow site
of Cu(111). Interaction between the Cu(111) substrate and BN leads to
redistribution of charge of BN sheet since the BN plane is a little rumpled.
Thus the gap opening of graphene can be understood based on the symmetry
breaking the A and B sublattice equivalence, which leads to the rehybridization
of the valence and conduction band states associated with the same Dirac point
(see Fig.3(b) and Fig.3(c)). This gap is essential for graphene-based
electronic devices for controlling conductivity.

Although the Dirac point of graphene is located in the energy gap of the
Cu(111) substrate at point K of reciprocal space, the electronic structure of
the graphene layer at the other points of momentum space is unknown, which
would have influence on conductivity. The K-resolved DOS, which is equivalent
to ARUPS experiment, of the graphene layer of the A2 structure is shown in
Fig.3(a). The band structure is almost the same as that of perfect graphene
except  the position of the Fermi-level (The energy gap cannot be seen clearly
at K which is due to the resolution in the calculated DOS). No other states
were observable in this graphene layer. Therefore, the electronic property of
the graphene layer shows features of a typical $n$-doped graphene. However, it
is noted that this occurs without an applied gate-voltage,\cite{gate3} and the
$n$-doped effect is completely induced by the substrate. Fig.3(b) shows the
precise K-resolved DOS around K of the carbon atom occupying the atop B site
and Fig.3(c) shows that of the carbon atom occupying the other sublattice.
Around the K point, the valence band and conduction band of the two sublattices
do not touch each other due to the symmetry breaking of the two sublattices.
They are different because of the rehybridization of the valence band and
conduction band states at the same Dirac point, as illustrated in Fig.~3(d).
For carbon over boron, the conduction bands degenerate while for carbon in the
other sublattice, the valence bands degenerate. The separation between the
conduction band peak of one carbon and the valence band peak of the other
carbon at K is about 0.18 eV which is in good agreement with  the value
determined from calculated band structure. The Fermi level is above this gap
due to substrate induced doping. The valence Dirac cone is populated with
electrons which have nearly zero effective-mass and very high mobility.

Similar calculations and analysis were carried out for structure A1, and its
electronic structure is nearly the same as that of A2. The only noticable
difference is the slightly small band gap, 0.15 eV, which is due to the
different small difference in atomic structures.

This $n$-doped graphene on the BN/Cu(111) substrate can be very useful in
realization of  graphene-based electronic devices. It was demonstrated recently
that $p$-type graphene can be easily realized experimentally by surface
modification with  electron acceptor molecule.\cite{pdoped} But a simple and
effective method to fabricate $p$-$n$ junction based on graphene is still
lacking. Based on results of present study, it would be possible to massive
produce $p$-$n$ junctions by  depositing electron acceptor molecules on certain
regions of graphene eptaxially grown on BN/Cu(111) substrate. Furthermore,
because the gap is just 0.5 eV below the Fermi-level, the Fermi-level can be
easily shifted inside the gap by adsorption of molecule or applying a gate
voltage to make graphene a real semiconductor. With a 0.2 eV band gap, graphene
would behave like a narrow-gap semiconductor, such as PbTe or PbSe. It may be
used as a material in infrared light emitting or detecting devices. One example
would be light emitting quantum dots, which have much more advantages than
conventional devices.

In conclusion, we have carried out a first principles investigation on graphene
grown on Cu(111) surface with a h-BN buffer layer. Electrons transfer from the
copper substrate to graphene through the BN buffer layer is predicted which
results in a $n$-doped graphene in the absense of a gate voltage. A gap opening
at the Dirac point just 0.5 eV below the Fermi-level is observed. The size of
the gap (0.2 eV) is comparable to that of a typical narrow gap semiconductor.
The Fermi-level can be easily shifted inside this gap to make graphene a
semiconductor. Considering that the gap size of graphene on h-BN or clean
Cu(111) is only a few meV,\cite{GrhBN} it would be possible to engineering the
gap of graphene through the tuning the interaction between the buffer layer and
the metal surface by choicing different metal surface or buffer layer.

\newpage
\begin{table}
\caption{Interlayer distances, as indicated in Fig.~1, of fully relaxed structures. All values are given in \AA.}
\begin{ruledtabular}
\begin{tabular}{ccccc}
           & $d_0$  & $d_1$ & $d_2$ & $d$  \\   \hline
    A1     & 3.036  & 2.485 & 1.990 & 2.024   \\
    A2     & 3.044  & 2.499 & 1.991 & 2.022  \\
    Cu(111)&        &       & 2.010 & 2.023\\
\end{tabular}
\end{ruledtabular}
\end{table}

\newpage
\begin{figure}[h]
\caption{The two most stable structures of graphene grown on Cu(111) with a h-BN buffer layer between them. A top view and a side view are presented in each case. The Cu atoms are represented using big grey spheres and their darkness increases with the distance from the surface. The carbon atoms are represented by small dark spheres. Nitrogen and boron atom are represented by medium sized blue and pink spheres respectively. $d_0$, $d_1$, $d_2$ and $d$ represent interlayer spacings and their values are given in Table I.}
\end{figure}

\begin{figure}[h]
\caption{
Band structure of (a) clean Cu(111); (b) h-BN on Cu(111); and (c) graphene on the BN/Cu(111). The Fermi level is indicated by dash line.}
\end{figure}

\begin{figure}[h]
\caption{
(a) K-resolved DOS of the graphene layer. (b) K-resolved DOS of carbon atop B along the path  indicated in (d). (c) Same as (b) but for carbon of the other sublattice. (d) Schematic diagram to show the band opening at Dirac point due to symmetry breaking between the two sublattice (A \& B).}
\end{figure}

\end{document}